\documentclass[a4paper,11pt]{article}
\usepackage{pos}

\title{Puzzle for the Vector Meson Threshold Photoproduction}
\ShortTitle{Puzzle...}

\author*[a]{Igor Strakovsky}

\affiliation[a]{Institute for Nuclear Studies, Department of Physics,\\
  The George Washington University, Washington, DC 20052, USA}

\emailAdd{igor@gwu.edu}

\abstract{High-statistics total cross sections for the vector meson photoproduction at the threshold: $\gamma p\to\omega p$ (from A2 at MAMI, ELPH, and CBELSA/TAPS), $\gamma p\to \phi p$ (from CLAS and LEPS), and $\gamma p\to J/\psi p$ (from GlueX) allow one to extract the absolute value of vector meson nucleon scattering length using Vector Meson Dominance (VMD) model. The ``young'' vector meson hypothesis may explain the fact that the obtained scattering length value for the nucleon $\phi$-meson compared to the typical hadron size of approximately $\sim1~\mathrm{fm}$ indicates that the proton is more transparent for the $\phi$-meson compared to the $\omega$-meson and is much less transparent than the $J/\psi$-meson. 
The extended analysis of $\Upsilon$-meson photoproduction using quasi-data from the QCD approach is in perfect agreement with the light-meson findings using experimental data.

Future high-quality experiments by EIC and EicC will have the opportunity to evaluate cases for $J/\psi$- and $\Upsilon$-mesons. It allows one to understand the dynamics of $c\bar{c}$ and $b\bar{b}$ production at the threshold. 
The ability of J-PARC to measure $\pi^- p\to \phi n$ and $\pi^-p\to J/\psi n$, which are free from the VMD model, is considered.}

\FullConference{%
  The 21st International Conference on Hadron Spectroscopy and Structure - HADRON2025 \\
  27-31 March, 2025 \\
  Toyonaka Campus, Osaka University, Osaka, Japan
}

\begin{document}
\maketitle

\section{Introduction}
\begin{figure*}[htb!]
\centering
{
    \includegraphics[width=0.4\textwidth,keepaspectratio]{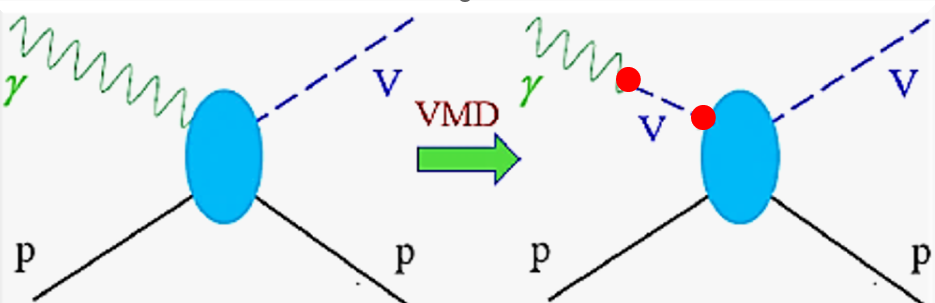}
}

\centerline{\parbox{0.9\textwidth}{
\caption[] {\protect\small
 Schematic diagrams of vector-meson photoproduction (left) and the VMD model (right) in the energy region at threshold experiments.
} 
\label{fig:fig0} } }
\end{figure*}

There are no vector meson (V) beams, so experiments using modern electromagnetic (EM) facilities attempt to access vector meson nucleon (VN) interactions via EM production reactions $ep\to e’Vp$. Some Vs can, compared to other mesons, be measured with very high precision. This comes from the fact that Vs have the same quantum numbers as the photon: $I^G(J^{PC}) = 0^-(1^{-~-})$. It allows us to apply a Vector Meson Dominance (VMD) model, assuming that a real photon can fluctuate into a virtual V, which subsequently scatters off a target nucleon (Fig.~\ref{fig:fig0})~\cite{Gell-Mann:1961jim, Kroll:1967it, Sakurai:1969jj}.
Let us focus on 4 Vs ($\omega$, $\phi$, $J/\psi(1S)$, and $\Upsilon(1S)$) from $q\bar{q}$ nonet, the widths of which are narrow enough to study meson photoproduction at threshold and where data and quasi-data are available. To avoid a broad width problem at threshold, we are not considering the $\rho$-meson case to determine VN scattering length (SL). Furthermore, we will ignore, for example, $\psi’(2S)$ due to the difference between the $1S$ and $2S$ states due to ``zero'' in radial wave functions (WFs). Unfortunately, we cannot go above Quarkonium or $\Upsilon$, whose quark content is $b\bar{b}$. The problem is that actually the Toponium ($T(1S)$), whose quark content is $t\bar{t}$, does not exist. It is due to a large mass of the Theta meson and the $t$-quark decays faster than the quarks form the Theta meson.

\section{Vector Meson Nucleon Scattering Length}
\begin{figure*}[htb!]
\centering
{
    \includegraphics[width=0.5\textwidth,keepaspectratio]{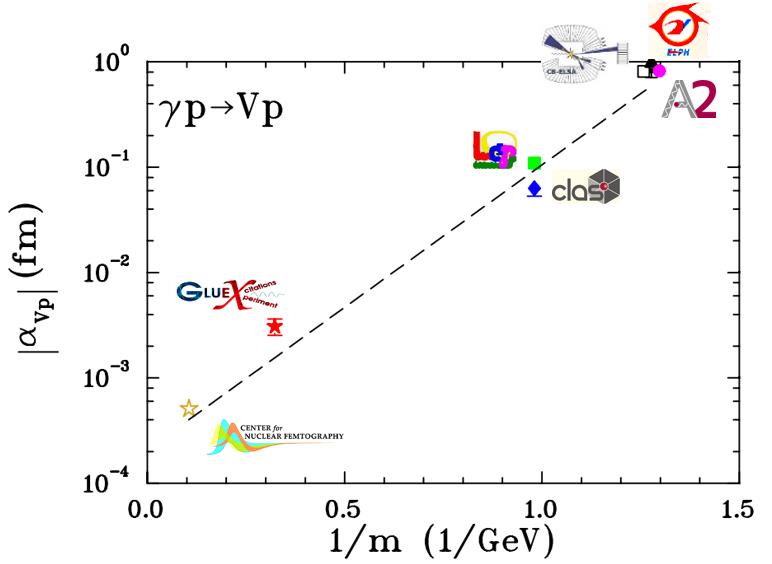}
}

\centerline{\parbox{0.9\textwidth}{
\caption[] {\protect\small
Comparison of the $|\alpha_{Vp}|$ SLs estimated from threshold V photoproduction on the proton target with VMD model contribution vs the inverse mass of the Vs. 
Input data for phenomenological analyses came from 
A2 at MAMI (magenta filled circle)~\cite{Strakovsky:2014wja}, 
ELPH (black filled triangle)~\cite{Ishikawa:2019rvz}, and 
CBELSA/TAPS (black open square)~\cite{CBELSATAPS:2015wwn} Collaborations for the $\omega$-meson;
CLAS (blue filled diamond)~\cite{Dey:2014tfa} and 
LEPS (green filled square)~\cite{LEPS:2005hax, Chang:2007fc} Collaborations for the $\phi$-meson; and
GlueX (red filled star)~\cite{GlueX:2019mkq} Collaboration for the $J/\psi$-meson; and quasi-data from Center for Nuclear Femtography (brown open star)~\cite{Guo:2021ibg} for the $\Upsilon$-meson.
Analyses results for
$\omega$-meson is given at Refs.~\cite{Strakovsky:2014wja, Ishikawa:2019rvz, Han:2022khg};
for $\phi$-meson is given at Refs.~\cite{Strakovsky:2020uqs, Han:2022khg};
for $J/\psi$-meson is given at Ref.~\cite{Strakovsky:2019bev}; and
for $\Upsilon$-meson is given at Ref.~\cite{Strakovsky:2021vyk}.
The black dashed line is hypothetical following $|\alpha_{Vp}| \propto 1/m_V$.
} 
\label{fig:fig1} } }
\end{figure*}

Due to the small size of ``young'' V vs ``old'' one, measured and predicted SL is very small. V created by the photon at the threshold, then most probably V is not completely formed and its radius is smaller than that of normal (``old'' ) V~\cite{Feinberg:1980yu}.  Therefore, a stronger suppression for the Vp interaction is observed (Fig.~\ref{fig:fig1}). $p\to V$ coupling $q\bar{q}$ is proportional to $\alpha_S$ and the separation of the corresponding quarks. This separation (with a zero approximation) is proportional to $1/m_V$, where $m_V$ is the mass of V.
\begin{table}[htb!]

\centering \protect\caption{{
The decay $\Gamma(V\to e^+e^-)$ from PDG2024~\cite{ParticleDataGroup:2024pth} (second column).
The third column showed the minimal momentum q$_{min}$ for V in photoproduction 
experiments on the proton target and the source of data.  
The fourth column showed Vp SLs, $|\alpha_{Vp}|$, and the sources of results.}
}

\vspace{2mm}
{%
\begin{tabular}{|c|c|c|c|}
\hline
Meson         & $\Gamma(V\to e^+e^-)$ & q$_{min}$ & $|\alpha_{Vp}|$     \tabularnewline
              &   (keV)               & (MeV/c)   & (fm)        \tabularnewline
\hline
$\omega(782)$ &  0.60$\pm$0.02        & 49~\cite{Strakovsky:2014wja} & 
0.82$\pm$0.03~\cite{Strakovsky:2014wja} \tabularnewline
            &                         &                              &
             0.97$\pm$0.16~\cite{Ishikawa:2019rvz} \tabularnewline
            &                         &                              &
             0.811$\pm$0.019~\cite{Han:2022khg}      \tabularnewline
$\phi(1020)$  &  1.27$\pm$0.04        & 216~\cite{Dey:2014tfa} & 
0.063$\pm$0.010~\cite{Strakovsky:2020uqs} \tabularnewline
            &                         &                              &
             0.109$\pm$0.008~\cite{Han:2022khg}      \tabularnewline
$J/\psi(1S)$  &  5.53$\pm$0.10        & 230~\cite{GlueX:2019mkq} & 
(3.08$\pm$0.55)$\times 10^{-3}$~\cite{Strakovsky:2019bev} \tabularnewline
$\Upsilon(1S)$&  1.340$\pm$0.018      & 521~\cite{Guo:2021ibg} & 
(0.51$\pm$0.03)$\times 10^{-3}$~\cite{Strakovsky:2021vyk} \tabularnewline
\hline
\end{tabular}} \label{tbl:tab1}
\end{table}

\begin{figure*}[htb!]
\centering
{
    \includegraphics[width=0.5\textwidth,keepaspectratio]{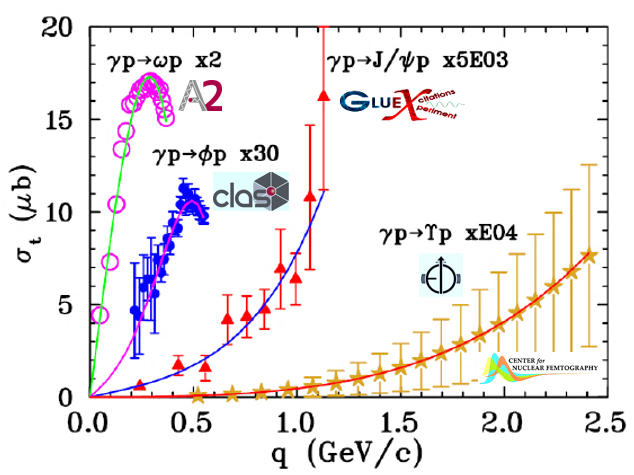}
}

\centerline{\parbox{0.9\textwidth}{
\caption[] {\protect\small
The total $\gamma p \to Vp$ cross section $\sigma_t$ derived from the A2 at MAMI (magenta open circles)~\cite{Strakovsky:2014wja}, CLAS (blue filled circles)~\cite{Strakovsky:2020uqs}, and GlueX (red filled triangles)~\cite{Strakovsky:2019bev}, data, and EIC/EicC (yellow
filled stars) quasi-data~\cite{Strakovsky:2021vyk} is shown as a function of the center of mass momentum $q$ of the final-state particles. The vertical (horizontal)
error bars represent the total uncertainties of the data summing statistical and systematic uncertainties in quadrature (energy binning which we did not use in our fits). Solid curves are the fits of the data 
.} 
\label{fig:fig2} } }
\end{figure*}

Small positive or negative VN SL may indicate a repulsive or attractive VN interaction if there is no VN bound state below experimental q$_{min}$ (here, 
q$_{min}$ is the V center-of-mass momentum), see Table~\ref{tbl:tab1}. For evaluation of the absolute value of VN SL, we apply the VMD model that links the near-threshold photoproduction cross sections of $\gamma p \to Vp$ and elastic $Vp \to Vp$. Finally, the absolute value of the VN SL can be expressed as a product of the kinematic factor motivated by the pure VMD of the EM ($V\to e^+e^-$ decay~\cite{ParticleDataGroup:2024pth} - see Table~\ref{tbl:tab1}) and the hadronic factor determined by the interaction of strong (hadronic) dynamics and EM (via fit of the total cross sections, $\sigma_t$, of the reactions $\gamma p\to Vp$ by a series of odd powers in $q$)~\cite{Strakovsky:2019bev, Strakovsky:2020uqs}.\footnote{To avoid theoretical uncertainties, we did not (i) determine the sign of SL, (ii) separate real and imaginary parts of SL, and (iii) extract spin 1/2 and 3/2 contributions.} Dramatic differences in hadronic factors as slopes of $\sigma_t$ at the threshold as a function of $q$ vary significantly from $\omega$ to $\phi$ to $J/\psi$ and to $\Upsilon$ (Fig.~\ref{fig:fig2}). Therefore, such a big difference in SL is determined mainly by the hadronic factor. The list of SLs (Fig.~\ref{fig:fig1}) is given in Table~\ref{tbl:tab1}.

A number of previous theoretical results (including potential approaches and LQCD calculations) gave much larger SLs. Most probably such large SL results from large distances in the tail of the van der Waals potential that in QCD should be killed by confinement~\cite{Dokshitzer:2023}.

There is no alternative to the VMD application to get VN SL from the vector meson photoproduction~\cite{Vainshtein:2020, Ryskin:2020}. A possible alternative is to develop a sophisticated, nonperturbative reaction theory that can explain $q\bar{q}$ scattering from hadron targets into vector meson final states.
To estimate theoretical uncertainty related to the VMD model, one refers to the estimation of the cross section of $J/\psi$ photoproduction in the peripheral model and finds strong energy dependence close to threshold because non-diagonal $\gamma p\to Vp$ and elastic $Vp\to Vp$ must have larger transfer momenta vs elastic scattering. It results in violation of VMD by a factor of 5~\cite{Boreskov:1976dj}. Independently, color factor for Charmonium is 1/9 while for open charm it is 8/9~\cite{Kopeliovich:2017jpy}.

In a recent study, the effect of the VMD assumption was studied in the formalism of Dyson-Schwinger equations, which one can consider as an alternative interpretation of the ``young age'' effect in another (more formal) language~\cite{Xu:2021mju}.

\section{Recent Results for Vector Mesons Nucleon SL}
One can see a high-energy experimental and LQCD activity addressed to the vector mesons ($\phi$ and $J/\psi$) nucleon SLs.

\subsection{High Energy Data for $\phi N$ SL}
Recently, the ALICE Collaboration has deduced spin-averaged $\phi N$ SL from the two-particle momentum correlation function~\cite{ALICE:2021cpv}. The attractive value is close to $\sim1~\mathrm{fm}$ ($\alpha_{\phi N} = [(0.85\pm 0.48) + i(0.16\pm 
0.19)]~\mathrm{fm}$.) In fact, ALICE is doing two-particle correlations of combined pairs $p\phi$ and $\bar{p}\phi$ measured in high multiplicity in $pp$ collisions at $W = 13~\mathrm{TeV}$. In addition, the final-state interaction (FSI) correlation C(k) depends on the production mechanism. Then, ALICE assumes that the protons and $\phi$-mesons are produced independently at the $\sim1~\mathrm{fm}$ distance. Another problem is that it is practically impossible to observe $p\phi$ (or any pV) correlation (with very small $p\phi$ energy, \textit{i.e.}, near threshold) at CERN (with ALICE or another detector).

The alternative analysis was reported by several members of the ALICE Collaboration for two models: $\alpha_{\phi p} = [0.272\pm i0.189]~\mathrm{fm}$ (pure theoretical) and $\alpha_{\phi p} = [(-0.034\pm 0.035) + i(0.57\pm 0.09)]~\mathrm{fm}$ (bootstrap model)~\cite{Feijoo:2024bvn}. Let us not discuss reasons for differences in results that came from the analyses of the same high-energy $W = 13~\mathrm{TeV}$ data and why the alternative analysis reported $ReSL\sim ImSL$, or even $ReSL << ImSL$, or why this treatment required a contribution from unknown $N^\ast(1700)$~\cite{ParticleDataGroup:2024pth}. Obviously, femtoscopic data (used in this analysis) are not sufficiently precise to provide new information for SL (besides, this femtoscopic data is affected by many other effects). Most probably, the radius of an event with $p\psi$ production is NOT equal to that measured via average $\pi\pi$ Bose-Einstein correlations~\cite{Schegelsky:2016gcy, Schegelsky:2016dzs, Schegelsky:2015sga}.

\subsection{LQCD results for $\phi N$ SL}
Using (2 + 1)-flavor lattice QCD simulations with nearly physical quark masses, HAL Collaboration has simulated the $N\phi$ scattering process for the spin 3/2 channel $\alpha_{\phi N}^{(3/2)} = [-1.43\pm 0.23^{+0.36}_{-0.06}]~\mathrm{fm}$~\cite{Lyu:2022imf}.  The results are compatible with the recent ALICE data~\cite{ALICE:2021cpv}.  Instead of the $\phi$ photoproduction process, the authors simulated the $\phi N$ elastic scattering reaction. $\phi N$ system is assumed to be ``on lattice'' and this result is but a ``numerical experiment.'' Using lattice calculations for the spin 3/2 $\phi N$ interaction by HAL Collaboration is used to constrain the spin 1/2 counterpart from the fit of the experimental correlation function $\phi p$ measured by ALICE: $\alpha_{\phi N}^{(1/2)} = [-(1.54^{+0.53}_{-0.53}{^{+0.16}_{-0.09}}) + i(0.00^{+0.34}_{-0.00}{^{+0.16}_{-0.00}})]~\mathrm{fm}$~\cite{Chizzali:2022pjd}. The corresponding SL is compatible with recent spin 3/2 results by the HAL~\cite{Lyu:2022imf} and ALICE~\cite{ALICE:2021cpv} results. The combination of HAL's spins 3/2 and 1/2 gives a huge $p\phi$ SL ($\alpha_{VN} = (2\alpha^{3/2}_{VN} + \alpha^{1/2}_{VN})/3$), which is much larger than the size of the hadron. Let us note, negative sign for both 3/2 and 1/2 components of $\pi N$ SL depends on the definition~\cite{Lyu:2025}.

\subsection{Model Dependent Analysis for $J\psi N$ SL}
\begin{figure*}[htb!]
\centering
{
    \includegraphics[width=0.35\textwidth,keepaspectratio]{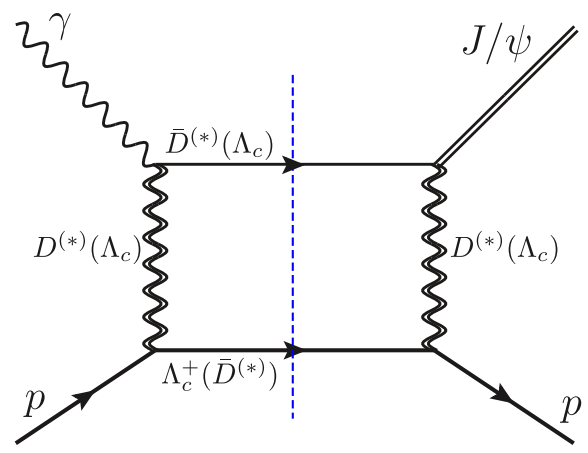}
}

\centerline{\parbox{0.9\textwidth}{
\caption[] {\protect\small
Feynman diagram for the proposed $cc$ mechanism. The dashed blue line pinpoints the open-charm intermediate state.
} 
\label{fig:fig3} } }
\end{figure*}

The suggested approach can be employed to evaluate the $J/\psi$-nucleon SLs, replacing the photon by a $J/\psi$-meson in Fig.~\ref{fig:fig3}. The results then appear to have the order of several units of $10^{-3}~\mathrm{fm}$, $\alpha_{J\psi N}^{(J=1/2)} = (0.2...3.1)\times 10^{-3}~\mathrm{fm}$ and $\alpha_{J\psi N}^{(J=3/2)} = (0.2...3.0)\times 10^{-3}~\mathrm{fm}$, where $J$ corresponds to the total angular momentum of the $J/\psi$-nucleon system~\cite{Du:2020bqj}. These numbers are comparable with our phenomenological estimation of the $J/\psi p$ SL from the GlueX data using the VMD model~\cite{Strakovsky:2019bev}.

\subsection{LQCD results for $J\psi N$ SL}
Recently, HAL Collaboration reported $\alpha_{J/\psi N}^{(1/2)} = [0.38\pm 0.04^{+0.00}_{-0.03}1~\mathrm{fm}$ and $\alpha_{J/\psi N}^{(3/2)} = [0.30\pm 0.02^{+0.00}_{-0.02}]~\mathrm{fm}$~\cite{Lyu:2025jjl}. Instead of $Vp\to Vp$ as they did for the $\phi$ case, they did $(c\bar{c})p\to (c\bar{c})p$. Different signs between $\phi N$ and $J/\psi N$ SLs in HAL's calculations cannot cause a problem because the sign depends on the definition~\cite{Lyu:2025}. But I do not see/know an explanation for the difference in $J/\psi$ SLs between HAL~\cite{Lyu:2025jjl} and our phenomenological results~\cite{Strakovsky:2019bev}.

\section{J-PARC Pion Induced Measurements}
There are no threshold total cross sections for $\pi^- p\to \phi n$ and $\pi^-p\to J/\psi n$ measurements, while theoretical extrapolation does not allow us to get an estimation for vector meson neutron SL (see, for instance, Refs.~\cite{Doring:2008sv, Kim:2016cxr}).
Let us note that following Ref.~\cite{Kim:2016cxr}, the cross section difference between $\pi^- p\to \phi n$ and $\pi^-p\to J/\psi n$ has a factor of $10^6$.

Actually, J-PARC is able to measure both $\pi^- p\to \phi n$~\cite{Ishikawa:2025} and $\pi^-p\to J/\psi n$~\cite{Noumi:2019} at the threshold. That is important to help solve a puzzle for VN SL, specifically, that analyses of these data are free from the VMD model. Crucial point for the $\pi^-p \to Vn$ case (as J-PARC will measure) is that the photon creates/produces $q\bar{q}$ pair at POINT while in the pion case we deal with two quarks separated one from another at the beginning.

Depending on the production mechanism for $J/\psi$, the polarization of $J/\psi$ could change. For example, some production mechanisms are expected to produce a largely unpolarized $J/\psi$ (such as inclusive production), while other mechanisms can produce a polarized $J/\psi$ (such as exclusive production or threshold production). For example, when the NuSea Collaboration measured the polarization of $J/\psi$ produced in the Fermilab $pA$ inclusive production at $800~\mathrm{GeV}$ beam energy, they found that $J/\psi$ is somewhat polarized, but much less polarized than the Drell-Yan di-leptons~\cite{NuSea:2000vgl, NuSea:2003fkm}. Many studies on the $J/\psi$ (or in general, quarkonium production) polarization have also been performed at collider energies. 

\section{Conclusion}
``Young'' hypothesis may explain the fact that the obtained SL value for the $\phi$ -meson nucleon compared to the typical hadron size of $1~\mathrm{fm}$ indicates that the proton is more transparent for the $\phi$-meson compared to the $\omega$ -meson and is much less transparent than $J/\psi$, and $\Upsilon$-meson.  Future high-quality experiments by EIC and EicC will have the opportunity to evaluate physics for $J/\psi$ and $\Upsilon$-mesons.  It allows us to understand the dynamics of $c\bar{c}$   and $b\bar{b}$ production at threshold. The ability of J-PARC to measure $\pi^-p\to \phi n$ and $\pi^-p\to J/\psi n$ at VMD-free thresholds is an important input in phenomenology (PWA). Polarized measurements are an important contribution for model-independent PWA.

\acknowledgments
I thank Misha Ryskin, Yuri Dokshitzer, Atsushi Hosaka, Takatsugu Ishikawa, Yan Lyu, Jen-Chieh Peng, Lubomir Pentchev, Sasha Titov, and Arkady Vainshtein for valuable comments and discussions.
This work was supported in part by the U.~S.~Department of Energy, Office of Science, Office of Nuclear Physics, under award No. DE--SC0016583. 



\begin{thebibliography}{99}
\bibitem{Gell-Mann:1961jim}
    M.~Gell-Mann and F.~Zachariasen,
    ``Form-factors and vector mesons,''
    Phys.\ Rev.\ \textbf{124}, 953 (1961).
\bibitem{Kroll:1967it}
    N.~M.~Kroll, T.~D.~Lee, and B.~Zumino,
    ``Neutral vector mesons and the hadronic electromagnetic current,''
    Phys.\ Rev.\ \textbf{157}, 1376 (1967).
\bibitem{Sakurai:1969jj}
    J.~J.~Sakurai, ``Currents and Mesons,'' 
    (The University of Chicago Press, Chicago, 1969).
\bibitem{Feinberg:1980yu}
  E.~L.~Feinberg,
  ``Hadron clusters and half dressed particles in Quantum Field Theory,''
  Usp.\ Fiz.\ Nauk\ \textbf{132}, 225 (1980) [Sov.\ Phys.\ Usp.\ \textbf{23}, 629 (1980)].
\bibitem{Strakovsky:2014wja}
    I.~I.~Strakovsky \textit{et al.} [A2 Collaboration at MAMI],
    ``Photoproduction of the \ensuremath{\omega} meson on the proton near threshold,''
    Phys.\ Rev.\ C\ \textbf{91}, 045207 (2015).
\bibitem{Ishikawa:2019rvz}
    T.~Ishikawa \textit{et al.}
    ``$\omega N$ scattering length from $\omega$ photoproduction on the proton near the threshold,''
    Phys.\ Rev.\ C\ \textbf{101}, 052201 (2020).
\bibitem{CBELSATAPS:2015wwn}
    F.~Dietz \textit{et al.} [CBELSA/TAPS Collaboration],
    ``Photoproduction of $\omega$ mesons off protons and neutrons,''
    Eur.\ Phys.\ J.\ A\ \textbf{51}, 6 (2015).
\bibitem{Dey:2014tfa}
    B.~Dey \textit{et al.} [CLAS Collaboration],
    ``Data analysis techniques, differential cross sections, and spin density matrix elements for the reaction $\gamma p \rightarrow \phi p$,''
    Phys. Rev. C \textbf{89}, 055208 (2014).
\bibitem{LEPS:2005hax}
    T.~Mibe \textit{et al.} [LEPS Collaboration],
    ``Diffractive phi-meson photoproduction on proton near threshold,''
    Phys.\ Rev.\ Lett.\ \textbf{95}, 182001 (2005).
\bibitem{Chang:2007fc}
    W.~C.~Chang, K.~Horie, S.~Shimizu, M.~Miyabe, D.~S.~Ahn, J.~K.~Ahn, H.~Akimune, Y.~Asano, S.~Date, H.~Ejiri \textit{et al.}
    ``Forward coherent phi-meson photoproduction from deuterons near threshold,''
    Phys.\ Lett.\ B\ \textbf{658}, 209 (2008).
\bibitem{GlueX:2019mkq}
    A.~Ali \textit{et al.} [GlueX Collaboration],
    ``First measurement of near-threshold J/\ensuremath{\psi} exclusive photoproduction off the proton,''
    Phys.\ Rev.\ Lett.\ \textbf{123}, 072001 (2019).
\bibitem{Guo:2021ibg}
    Y.~Guo, X.~Ji, and Y.~Liu,
    ``QCD analysis of near-threshold photon-proton production of heavy quarkonium,''
    Phys.\ Rev.\ D\ \textbf{103}, 096010 92021).
\bibitem{Han:2022khg}
    C.~Han, W.~Kou, R.~Wang, and X.~Chen,
    ``Extraction of \ensuremath{\omega}n, \ensuremath{\omega}p, and \ensuremath{\phi}N scattering lengths from \ensuremath{\omega} and \ensuremath{\phi} differential photoproduction cross sections~on a deuterium target,''
    Phys.\ Rev.\ C\ \textbf{107}, 015204 (2023).
\bibitem{Strakovsky:2020uqs}
    I.~I.~Strakovsky, L.~Pentchev, and A.~Titov,
    ``Comparative analysis of $\omega p$, $\phi p$, and $J/\psi p$ scattering lengths from A2, CLAS, and GlueX threshold measurements,''
    Phys.\ Rev.\ C\ \textbf{101}, 045201 (2020).
\bibitem{Strakovsky:2019bev}
    I.~Strakovsky, D.~Epifanov, and L.~Pentchev,
    ``J/$\psi$p scattering length from GlueX threshold measurements,''
    Phys.\ Rev.\ C\ \textbf{101}, 042201 (2020).
\bibitem{Strakovsky:2021vyk}
    I.~I.~Strakovsky, W.~J.~Briscoe, L.~Pentchev, and A.~Schmidt,
    ``Threshold Upsilon-meson photoproduction at the EIC and EicC,''
    Phys.\ Rev.\ D\ \textbf{104}, 074028 (2021).
\bibitem{ParticleDataGroup:2024pth}
    S.~Navas \textit{et al.} [Particle Data Group],
    ``Review of Particle Physics,''
    Phys.\ Rev.\ D\ \textbf{110}, 030001 (2024).
\bibitem{Dokshitzer:2023}
    Yu.~L.~Dokshitzer (private communication).
\bibitem{Vainshtein:2020}
    A.~Vainshtein (private communication).
\bibitem{Ryskin:2020}
    M.~Ryskin (private communication).
\bibitem{Boreskov:1976dj}
    K.~G.~Boreskov and B.~L.~Ioffe,
    ``$J/\psi$ meson photoproduction in the peripheral model,''
    Yad.\ Fiz.\ \textbf{25}, 806 (1977)
    [Sov.\ J.\ Nucl.\ Phys.\ \textbf{25}, 331 (1977)].
\bibitem{Kopeliovich:2017jpy}
    B.~Z.~Kopeliovich, I.~Schmidt, and M.~Siddikov,
    ``Suppression versus enhancement of heavy quarkonia in pA collisions,''
    Phys.\ Rev.\ C\ \textbf{95}, 065203 (2017).
\bibitem{Xu:2021mju}
    Y.~Z.~Xu, S.~Chen, Z.~Q.~Yao, D.~Binosi, Z.~F.~Cui, and C.~D.~Roberts,
    ``Vector-meson production and vector meson dominance,''
    Eur.\ Phys.\ J.\ C\ \textbf{81}, 895 (2021).
\bibitem{ALICE:2021cpv}
    S.~Acharya \textit{et al.} [ALICE Collaboration],
    ``Experimental evidence for an attractive p-$\phi$ interaction,''
    Phys.\ Rev.\ Lett.\ \textbf{127}, 172301 (2021).
\bibitem{Feijoo:2024bvn}
    A.~Feijoo, M.~Korwieser, and L.~Fabbietti,
    ``Relevance of the coupled channels in the \ensuremath{\phi}p and \ensuremath{\rho}0p correlation functions,''
    Phys.\ Rev.\ D\ \textbf{111}, 014009 (2025).
\bibitem{Schegelsky:2016gcy}
    V.~A.~Schegelsky and M.~G.~Ryskin,
    ``Small size sources of secondaries observed in pp-collisions via Bose-Einstein correlations at the LHC ATLAS experiment,''
    [arXiv:1608.05218 [hep-ph]].
\bibitem{Schegelsky:2016dzs}
    V.~A.~Schegelsky and M.~G.~Ryskin,
    ``Multiparticle production: an old-fashioned view,''
    [arXiv:1604.01189 [hep-ph]].
\bibitem{Schegelsky:2015sga}
    V.~A.~Schegelsky and M.~G.~Ryskin,
    ``Bose-Einstein correlation to measure the size of event of different types,''
    [arXiv:1506.03718 [hep-ph]].
\bibitem{Lyu:2022imf}
    Y.~Lyu, T.~Doi, T.~Hatsuda, Y.~Ikeda, J.~Meng, K.~Sasaki, and T.~Sugiura,
    ``Attractive N-\ensuremath{\phi} interaction and two-pion tail from lattice QCD near physical point,''
    Phys.\ Rev.\ D\ \textbf{106}, 074507 (2022).
\bibitem{Chizzali:2022pjd}
    E.~Chizzali, Y.~Kamiya, R.~Del Grande, T.~Doi, L.~Fabbietti, T.~Hatsuda, and Y.~Lyu,
    ``Indication of a p\textendash{}\ensuremath{\phi} bound state from a correlation function analysis,''
    Phys.\ Lett.\ B\ \textbf{848}, 138358 (2024).
\bibitem{Lyu:2025}
    Y.~Lyu (private communication).
\bibitem{Du:2020bqj}
    M.~L.~Du, V.~Baru, F.~K.~Guo, C.~Hanhart, U.~G.~Mei\ss{}ner, A.~Nefediev, and I.~Strakovsky,
    ``Deciphering the mechanism of near-threshold $J/\psi$ photoproduction,''
    Eur.\ Phys.\ J.\ C\ \textbf{80}, 1053 (2020).
\bibitem{Lyu:2025jjl}
    Y.~Lyu, T.~Doi, T.~Hatsuda, and T.~Sugiura,
    ``$NJ/\psi$ and $N\eta_c$ interactions from lattice QCD,''
    PoS \textbf{LATTICE2024}, 103 (2025).
\bibitem{Doring:2008sv}
    M.~Doring, E.~Oset, and B.~S.~Zou,
    ``The Role of the N*(1535) resonance and the $\pi^- p \to KY$ amplitudes in the OZI forbidden $\pi^- N \to \phi N$ reaction,''
    Phys.\ Rev.\ C\ \textbf{78}, 025207 (2008).
\bibitem{Kim:2016cxr}
    S.~H.~Kim, H.~C.~Kim, and A.~Hosaka,
    ``Heavy pentaquark states $P_c(4380)$ and $P_c(4450)$ in the $J/\psi$ production induced by pion beams off the nucleon,''
    Phys.\ Lett.\ B\ \textbf{763}, 358 (2016).
\bibitem{Ishikawa:2025}
    T.~Ishikawa \textit{et al.} [E95 Collaboration], ``Pion-induced phi-meson production on the proton,'' J-PARC Experimental Proposal E95 (2022); 
    https://j-parc.jp/researcher/Hadron/en/pac$_-$2208/pdf/P95$_-$2022-19.pdf .
\bibitem{Noumi:2019}
    W.~C.~Chang, H.~Noumi, and S.~Sawada,
    ``Studying Generalized Parton Distributions with exclusive Drell-Yan process at 
    J-PARC,'' J-PARC Letter-of-Intent (2019);
    https://j-parc.jp/researcher/Hadron/en/pac$_-$1901/pdf/LoI$_-$2019-07.pdf .
\bibitem{NuSea:2000vgl}
    C.~N.~Brown \textit{et al.} [NuSea Collaboration],
    ``Observation of polarization in bottomonium production at $\sqrt{s}$ = 
    38.8-GeV,''
    Phys.\ Rev.\ Lett.\ \textbf{86}, 2529 (2001).
\bibitem{NuSea:2003fkm}
    T.~H.~Chang \textit{et al.} [NuSea Collaboration],
    ``$J/\psi$ polarization in 800-GeV $p$ Cu interactions,''
    Phys.\ Rev.\ Lett.\ \textbf{91}, 211801 (2003).
\end{thebibliography}
\end{document}